\documentclass[aps,prb,showpacs,amsfonts,twocolumn]{revtex4}
\usepackage[dvips]{graphicx,color}

\usepackage{shortvrb}
\MakeShortVerb{\!} 

\newcommand{\LVSO}{Li$_2$VOSiO$_4$}

\begin{document}
\title{
Determination  of   the   exchange  energies    in   \LVSO~   from    a
high-temperature series analysis  of  the square  lattice  $J_1$-$J_2$
Heisenberg model.}

\author{G.~Misguich}
\email{gmisguich@cea.fr}
\affiliation{
Service de Physique Th{\'e}orique, CEA/DSM/SPhT -- URA  2306 of CNRS.\\
CEA/Saclay, 91191 Gif-sur-Yvette C\'edex, FRANCE}

\author{B.~Bernu}
\email{bernu@lptl.jussieu.fr}
\affiliation{
Laboratoire de Physique
Th{\'e}orique des Liquides, Universit{\'e} Paris-VI -- URA 765 of CNRS.\\
75252 Paris C\'edex, FRANCE.
}

\author{L.~Pierre}
\email{lpierre@lptl.jussieu.fr}
\affiliation{
Universit\'e Paris-X -- UMR 7600 of CNRS\\
92001 Nanterre C\'edex, FRANCE.}

\bibliographystyle{prsty}


\begin{abstract}

We  present   a high-temperature   expansion   (HTE)  of the  magnetic
susceptibility and specific heat  data of Melzi~{\it  et al.} on \LVSO
[Phys.   Rev.   B {\bf 64},  024409 (2001)].   The data  are very well
reproduced by  the $J_1$-$J_2$ Heisenberg  model on the square lattice
with  exchange energies $J_1=1.25\pm0.5$~K and $J_2=5.95\pm0.2$~K. The
maximum of the specific  heat $C_v^{\rm max}(T_{\rm max})$ is obtained
as a function $J_2/J_1$ from an improved method based on HTE.

\end{abstract}

\maketitle

\section{Introduction}

The   vanadium    oxide       \LVSO~is  a    quasi     two-dimensional
magnet~\cite{millet98}   which    is     well  described   by      the
spin-$\frac{1}{2}$ Heisenberg      model            on  the     square
lattice with first ($J_1$) and second ($J_2$) neighbors
interactions :
\begin{equation}
	H=J_1\sum_{<i,j>} \vec{S}_i\cdot\vec{S}_j
		+ J_2\sum_{<<i,j>>} \vec{S}_i\cdot\vec{S}_j
	\label{eq:H}
\end{equation}

In their   NMR experiments Melzi~{\it   et al.}~\cite{melzi00}   found
evidence  for a $Q=(\pi,0)$ or $(0,\pi)$  ordering  in this system, as
expected in  the classical  spin  model for  $J_2>0.5J_1$ (order  from
disorder)  and  also as  expected in  the spin-$\frac{1}{2}$  case for
$J_2{\gtrsim}0.6J_1$.~\cite{j1j2} From  the high temperature  behavior
of  $\chi(T)$   the    Curie-Weiss temperature   $\Theta=J_1+J_2$  was
estimated~\cite{melzi01} to be $\Theta\simeq8.2\pm1$~K. Combining this
information with the position of  the maximum of  the specific heat of
the  system  (and comparing  it to exact   diagonalization data for 16
spins),  Melzi   {\it et  al.}~\cite{melzi01}  estimated the  ratio of
exchange energies to be $J_2/J_1\simeq 1.1$.

In  two  recent papers,  Rosner  {\it et al.}~\cite{rosner02,rosner03}
considered the    determination of $J_1$ and  $J_2$   in  \LVSO.  They
performed  band-structure  calculations  (local-density approximation)
and found $J_1+J_2\simeq 9.5 \pm 1.5$~K  and $J_2/J_1\simeq 12$.  They
also   computed the high-temperature  expansion (HTE)  of the magnetic
susceptibility   and of the specific    heat  for the Hamiltonian   of
Eq.~\ref{eq:H}.  The $\chi(T)$ and $C_v(T)$ obtained from these series
at  $J_2/J_1\simeq    12$  are   in reasonable     agreement  with the
experimental  data  for  $\chi(T)$   and $C_v(T)$.   Rosner  {\it   et
al.}\cite{rosner02} did not  use the HTE  to fit the  data in the high
temperature region ($T>20$~K) because they could not  find a Curie law
behavior.   Melzi~{\it   et    al.}    indeed had    to     subtract a
temperature-independent constant to  recover a $1/T$ behavior at  high
temperature.   This  term is   due   to  the Van-Vleck   paramagnetism
(spin-orbit coupling  effect)  and  its magnitude   is  typical for  a
$V^{4+}$   ion  in a pyramidal    environment.~\cite{melzi01} It is of
course   crucial to  exploit   the  high   temperature region of   the
experimental data in order to take full advantage of the HTE.  In this
work we  show  that  the  HTE  for  the susceptibility  allows  us  to
determine the exchange energies in \LVSO.  We show that this method is
an  unbiased way of determining   the  microscopic parameters of  this
model.   As a   result we find   a ratio  $J_2/J_1\simeq 5$  which  is
significantly    different    from      the  value    $12$   predicted
previously.\cite{rosner02,rosner03} We  checked    this result by    a
calculation of  the  specific heat.  For this   purpose we employed  a
recent  technique~\cite{bm01} which  extends   the convergence  of the
series to  low  temperatures by  using two  constraints related to the
total entropy of the system and to its ground-state energy.

\section{High temperature series and extrapolation}

The series for $\chi(T)$ and $C_v(T)$ are computed in the standard way
by a  cluster expansion method to  order $1/T^{11}$.  Each coefficient
is a rational  number and is computed  exactly  (series available upon
request).  These    series      were  computed      independently   in
Ref.~\onlinecite{rosner03} to order $1/T^{10}$  and our results  agree
with their.

\section{Fit of the susceptibility}

\begin{figure}
	\includegraphics[width=6.7cm]{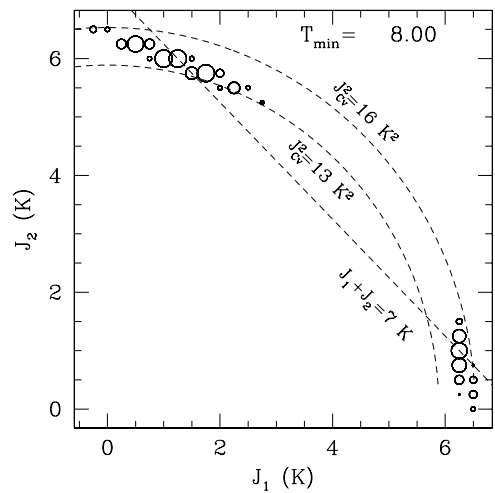}

	\caption{Quality  of the susceptibility fit   as a function of
	$J_1$   and  $J_2$.     The biggest  circles   correspond best
	fits.  Two  regions  around   $J_2\sim6$~K   and  $J_1\sim1$~K
	(resp. $J_2\sim1$~K and $J_1\sim6$~K)  appear to be compatible
	with the   data on  \LVSO.    The radius of the   circles  are
	proportional to   $r=8\Sigma^2_{\rm     min}-\Sigma^2$   where
	$\Sigma^2$ is   the   mean  square difference    between   the
	theoretical curve  and  experimental  data  and $\Sigma^2_{\rm
	min}$  corresponds   to   the  optimal  (smallest    value  of
	$\Sigma^2$).  Parameters which give $r<0$ are not represented.
	Dashed circles : $J_{C_v}^2=14.5\pm1.5$~K$^2$. The dashed line
	corresponds  to   a  Curie-Weiss temperature   $J_1+J_2=7$~K.} 
	\label{fig:fit2}

\end{figure}

In the first  step we  fit the  susceptibility data  of Melzi {\it  et
al.}~\cite{melzi00} by the following procedure:
\begin{itemize}
\item {\bf Temperature range of the fit}.
The  experimental  data are   fitted over  a fixed  temperature  range
$\left[T_{\rm min},T_{\rm  max}\right]$ where  $T_{\rm max}=300$~K  is
the     highest     available        temperature     available      in
Ref.~\onlinecite{melzi00} and      $T_{\rm     min}$  is    determined
self-consistently so that the susceptibilities calculated from the HTE
are well converged  down  to $T_{\rm  min}$ when  $J_1$ and  $J_2$ are
close to  the optimal parameters.  A rather  high $T_{\rm min}=8$~K is
chosen for the first global scan  of Fig.~\ref{fig:fit2}; this insures
that the calculated $\chi(T)$ is  well converged in the whole interval
whatever $(J_1,J_2)$.  Once the  relevant region in  $J_1$-$J_2$ space
is identified, $T_{\rm min}$ is lowered down to $T_{\rm min}=5.5$~K, a
temperature  still  above    that  where   the  various   approximants
approximants (Pad\'e) start to differ from each other.

\item {\bf Scan over $J_1$ and $J_2$}.
For each set of couplings, the HTE for $\chi(T)$ (including $1/T^{11}$
order)  is   extrapolated by all  possible   Pad\'e approximants.  The
rational fractions  which  have   (spurious) poles in    the  interval
$\left[T_{\rm min},T_{\rm max}\right]$ are discarded.

\item {\bf Measure of the error}.
The    experimental     data    $\chi^{exp}$    are      compared   to
$\chi_{VV}+C_0\chi^{th}(T)$  where     is $\chi_{VV}$         is     a
temperature-independent  Van-Vleck    term  and   $C_0$  the     Curie
constant.  For  each  value of $(J_1,J_2)$  and   for  each Pad\'e the
parameters $\chi_{VV}$  and  $C_0$ are determined   to insure the best
possible fit.  The quality of  the fit is measured  in  a standard way
through
$\Sigma^2=\sum_{T_{\rm min}\leq T_i \leq T_{\rm max}}{
	\left[C_0\chi(T_i)+\chi_{VV}-\chi^{exp}(T_i)\right]^2
	}$.
\end{itemize}

\begin{figure}
	\includegraphics[width=6.7cm]{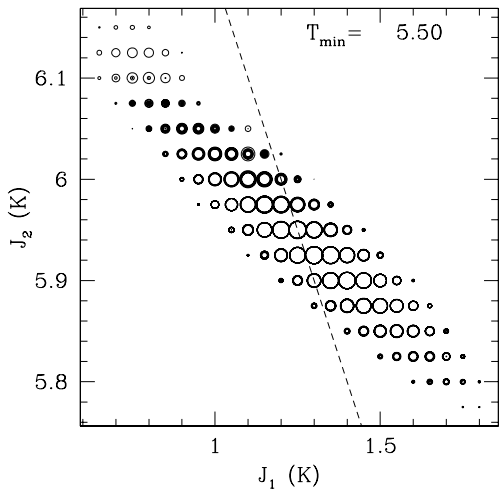}

	\caption{Quality of  the  susceptibility fit as a  function of
	$J_1$  and $J_2$.  Same  as Fig.~\ref{fig:fit2}.  When several
	concentric  circles are visible,  some Pad\'e approximants are
	significantly  different from  the  others at low temperature.
	The    points   fall       in the  rectangle       defined  by
	$J_1+3.25J_2=20.59\pm 0.11$~K and  $J_2-3.25J_1=1.9\pm 1.8$~K.
	}  \label{fig:fit0}
\end{figure}

\begin{figure}
	\includegraphics[width=6.7cm]{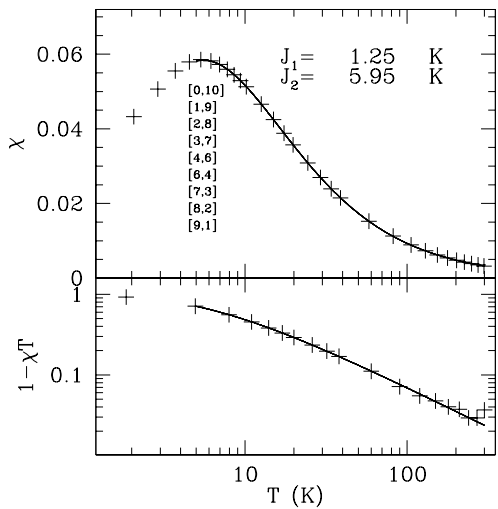}

	\caption{Top: Experimental susceptibility   (crosses) compared
	to the  HTE  calculation (full  lines).  The  different Pad\'e
	approximants  (degrees     $[u,d]$    indicated)    are    not
	distinguishable at  the scale of  the figure for $T>7$~K.  The
	best fit is obtained  for: Curie constant $C_0=0.333$ emu/mol,
	Van-Vleck                                       susceptibility
	$\chi_{VV}=4.014\times10^{-4}$emu/mol.  Bottom: Deviation from
	the Curie law (same data as top panel).  } \label{fig:chi}
\end{figure}

The figure~\ref{fig:fit2} shows  that   two regions of  the  parameter
space  are compatible   with the  susceptibility  of the   $J_1$-$J_2$
model. The  points  where $\Sigma^2$ is  larger  than eight times  its
smallest value (fit of poor quality) are not  displayed.  It turns out
that the Pad\'e  approximants to the  susceptibility  converge down to
$T\simeq 5.5$~K  in the two ``patches''   of Fig.~\ref{fig:fit2}.  For
this reason $T_{\rm min}$ can be lowered down to $5.5$~K to refine the
scans.  Although   both regions give fits   of comparable quality, the
existence of $(\pi,0)$  magnetic   ordering in  \LVSO~as well as   the
results of Ref.~\onlinecite{rosner02}   make the  $J_1>J_2$   scenario
extremely unlikely.  For this reason we will only  focus on the region
$J_1{\simeq}1.25$~K   and  $J_2{\simeq}6$~K  in the   following.   The
refined   scan is  shown  in  Fig.~\ref{fig:fit0}. The  susceptibility
obtained from  the HTE is   compared with the experimental  results in
Fig.~\ref{fig:chi}.  The Van-Vleck  susceptibility  and Curie constant
are parameters  of the  fit and  the  optimal  values are  in complete
agreement with   Ref.~\onlinecite{melzi01}.  In order to  estimate the
quality of  the  fit in the  high-temperature region  where the HTE is
almost exact, we subtracted the $1/T$  Curie term from the theoretical
curve  as well as  from the experimental  data. Even  when the leading
Curie term is subtracted (bottom of Fig.~\ref{fig:chi}), the deviation
of the  Pad\'e  approximants  from  the  experimental  results remains
within experimental error bars.

\section{Specific heat}

\begin{figure}
	\includegraphics[width=6.7cm]{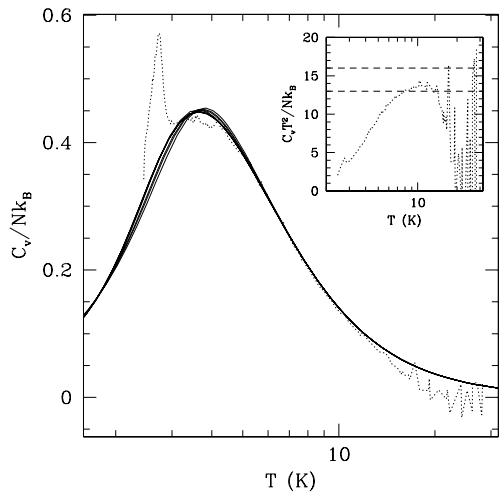}

	\caption{Specific  heat.   Full  lines: theoretical prediction
	for  $J_1=1.25$~K  and   $J_2=5.95$~K.   The  different  lines
	correspond  to   different    Pad\'es     approximants.  Their
	dispersion is  a measure  of  the  error  bars.  Dotted  line:
	experimental  data.     Inset:    experimental   specific heat
	multiplied by $T^2$.  We estimate that $C_vT^2/Nk_B\to 14.5\pm
	1.5$~$K^2$  (horizontal  lines).    The  agreement between the
	calculated $C_v$ and  the   experimental data can easily    be
	improved (data not shown)   at high temperatures  by  changing
	slightly the  $T^3$ contribution (phonons) that was subtracted
	from the raw data.~\cite{melzi01}} \label{fig:cv}
\end{figure}

At  high temperature,  the  specific  heat of  the $J_1$-$J_2$  model
behaves as:
\begin{eqnarray}
	C_v/(Nk_B)\sim \left(\frac{J_{C_v}}{T}\right)^2
		=\frac{3}{8}\left(J_1^2+J_2^2\right) \frac{1}{T^2}
\end{eqnarray}
Although the  specific  heat  data   are not  very  accurate  at  high
temperatures due to the subtraction of the phonon contribution, a plot
of  $C_vT^2$    shows    (see inset   of     Fig.~\ref{fig:cv})   that
$J_{C_v}^2{\simeq}14.5\pm1.5$~K$^2$.  This   constrains  the values of
$J_1$  and  $J_2$   to  lie   between  the    two  large   circles  in
Fig.~\ref{fig:fit2},   which is  compatible  with the  independent fit
performed on the susceptibility.

The  specific heat  is   obtained  from  the    HTE as   described  in
Ref.~\onlinecite{bm01} and will  not be reviewed  here. In addition to
the series   itself,  this method  requires   the knowledge   of three
quantities:
\begin{itemize}
	\item The total entropy, that is $\ln(2)$ per site. This means
	$\int_0^\infty C_v(T)/TdT=Nk_B\ln(2)$.

	\item     Ground-state  energy   per    site   $e_0$    of the
	Hamiltonian.  Since the  energy of  Eq.~\ref{eq:H}  is zero at
	infinite temperature we have $\int_0^\infty C_v(T)dT=-Ne_0$.

	\item The low-temperature behavior of  $C_v(T)$. It behaves as
	$\sim  T^2$    (resp.   $\sim  \exp(-\Delta/T)$)    when   the
	ground-state is N\'eel long-ranged ordered (resp. gapped).
\end{itemize}

The   ground-state   energy    of   the     first-neighbor  Heisenberg
antiferromagnet  on the square lattice  is known  very accurately from
quantum Monte  Carlo simulations~\cite{EJ1}:  $e_0=-0.6694 J_1$.  This
result also  applies  to the infinite-$J_2$   limit of the $J_1$-$J_2$
model.   In  the  frustrated  case  the ground-state energy   has been
computed by different  numerical techniques, such  as zero-temperature
series  expansion.\cite{swho99} From  these   result we constructed  a
simple ansatz which interpolates between the pure-$J_1$ and pure-$J_2$
models :
\begin{eqnarray}
	e_0=\alpha_1J_1+\alpha_2J_2
	-\sqrt{\left(
		\alpha_{11}J_1^2+\alpha_{22}J_2^2+\alpha_{12}J_1J_2
	\right)}
	\label{eq:e}
\end{eqnarray}
with  $\alpha_1=-0.3135$, $\alpha_2=-0.1207$,    $\alpha_{11}=0.1267$,
$\alpha_{22}=0.3011$   and $\alpha_{12}=-0.3722$.    The    comparison
between the different   approximants  for  the  specific heat    (with
$C_v\sim T^2$) at  $J_1=1.25$~K and $J_2=5.95$~K  and the experimental
data  is  shown Fig.~\ref{fig:cv}.   Unlike  usual Pad\'e approximants
directly constructed on   the   HTE   of   the  specific  heat,    our
procedure~\cite{bm01} leads to accurate results (with a relative error
smaller than a few percents) down to  zero temperature.  The agreement
with  the experimental data is  very good for $T>4.5$~K (no adjustable
parameter).  On  the  other hand  we  checked  that the  flat  maximum
observed in the experiment between 3~K and 4~K can hardly be accounted
for by the $J_1$-$J_2$ Heisenberg model, whatever the couplings.  This
is a  serious indication of a  structural distortion (above $T_c\simeq
2.8$~K corresponding to the (3D) magnetic  ordering of the system), as
suggested by Melzi~{\it et al}~\cite{melzi01} from the analysis of the
NMR spectra and   by Becca and  Mila~\cite{bm02}  from the theoretical
point of view.

When one of the $J$'s is much larger than the other, the specific heat
and the magnetic susceptibility are found to be nearly symmetric under
the  exchange  $J_1\leftrightarrow  J_2$  (at  least  at   no  too low
temperatures). Thus this analysis cannot strictly discriminate between
the  two patches of  Fig.~\ref{fig:fit2}.  Nevertheless $J_2>J_1$ gives
slightly better fits.

\begin{figure}
	\includegraphics[width=7.2cm]{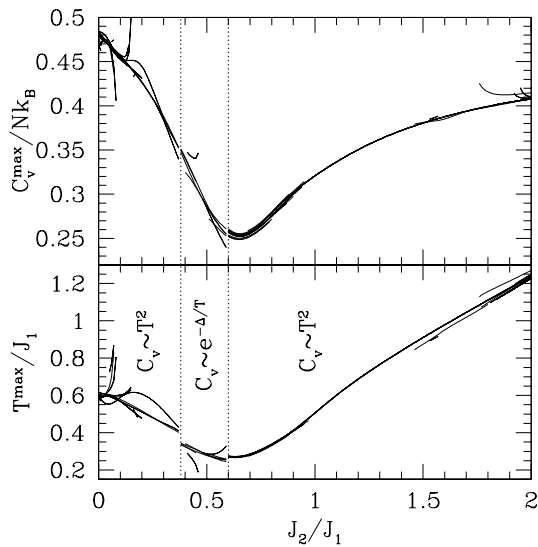}

	\caption{Maximum of the specific  heat and temperature of that
	maximum  as   functions of   $J_2/J_1$.  The  different curves
	correspond  to  the different  Pad\'e   approximants  at order
	$1/T^{11}$. The  regions where the model is  assumed to have a
	$\sim  T^2$  (resp.  $\sim  e^{-\Delta/T}$) specific heat  are
	indicated.} \label{fig:cvtmax}
\end{figure}

$C_v(T)$ for arbitrary $J_2/J_1$. --- The specific heat is computed by
the method above in a large range of coupling $J_2/J_1$.  The behavior
of $C_v$ at  low temperatures is assumed  to be: i) $C_v{\sim}T^2$ for
$J_2/J_1{\lesssim}0.38$ (corresponding  to an ordered antiferromagnet)
ii) $C_v\sim\exp(-\Delta/T)$ for   $0.38{\lesssim}J_2/J_1\lesssim 0.6$
(gapped   region~\cite{j1j2})   and    iii) $C_v\sim     T^2$   for  $
0.6{\lesssim}J_2/J_1  $ (ordered  antiferromagnet).  The  ground-state
energy is approximated by Eq.~\ref{eq:e}.   The results are summarized
in Fig.~\ref{fig:cvtmax} where  the  the value $C_v^{\rm max}$  of the
maximum  of  the specific heat and   the temperature  $T^{\rm max}$ at
which $C_v$ reaches its maximum are  displayed.  The dispersion of the
different approximants is  again   an  estimate of  the error.     The
discontinuities in the  curves  are due to the   fact that the  Pad\'e
approximants  which develop  zeros or   poles in  the  physical energy
interval cannot   be considered.\cite{bm01} The  results for $C_v^{\rm
max}$ and $T^{\rm max}$  represent a significant improvement over  the
estimate made in Ref.~\onlinecite{melzi01}.  The shape of the specific
heat   appears  to   be   relatively  independent  of  $J_2/J_1$  when
$J_2\gtrsim2J_1$ but some interesting structure appear in the strongly
frustrated region around  $J_2\simeq0.5J_1$.   The existence of a  low
$C_v^{\rm max}$  is indeed compensated  by a  small $T^{\rm  max}$  to
conserve the  total  entropy.   This shift  of   the entropy to  lower
temperatures and  lowed energies is indeed   expected close to quantum
phase transitions and in frustrated systems in general.

\section{Conclusions}

We   have computed  the HTE  for   the uniform  susceptibility of  the
$J_1$-$J_2$   Heisenberg model  on  the  square  lattice   up to order
$1/T^{11}$  and  obtained $J_1=1.25\pm0.5$~K and $J_2=5.95\pm0.2$~K by
fitting the experimental data of \LVSO.  The HTE for the specific heat
to the same order  and an improved  method  allowed us to  extrapolate
$C_v(T)$  down   to  $T=0$.   A   good  agreement is found   with  the
experimental    data of  \LVSO    for  $T\gtrsim4.5$~K.   Below   that
temperature  the specific  heat slightly differs   from our results on
$J_1$-$J_2$ Heisenberg model,   which is consistent  with  the lattice
distortion  transition  observed by  Melzi~{\it et al.}\cite{melzi01}
Chandra, Coleman  and   Larkin~\cite{ccl90}  predicted  an  Ising-like
finite temperature  phase  transition  in  the classical  and  quantum
$J_1$-$J_2$ Heisenberg models.  We could not find such a transition in
the present calculations  on  the quantum model   but some work is  in
progress to investigate thoroughly this question.~\cite{mila}

Acknowledgments.--- It  is  a  pleasure  to thank  Roberto   Melzi and
Philippe  Mendels for fruitful  discussions concerning the analysis of
the experimental data.


\end{document}